\documentclass[12pt,amsfonts,draft]{iopart}
\usepackage{amsfonts}
\begin{document}

\def\MR{\mathbb R}
\def\MC{\mathbb C}
\def\MN{\mathbb N}
\def\MZ{\mathbb Z}
\def\ID{\bf 1}
\def\ad{{\rm ad}}
\def\Tr{{\rm Tr}}
\def\sgn{{\rm sgn}}
\def\Im{\mathfrak{Im}}

\hbox to \hsize{\hss DAMTP-2004-93}

\hbox to \hsize{\hss hep-th/0409182}
\title[String quantization: Fock vs.\ LQG Representations]{String
  quantization:\\ Fock vs.\ LQG Representations} 
\author{Robert C. Helling, Giuseppe Policastro}
\address{DAMTP, Cambridge University, Cambridge CB3 0WA, UK}
\ead{helling@atdotde.de, g.policastro@damtp.cam.ac.uk}
\begin{abstract}
%%   We contrast the usual textbook quantization of the bosonic string
%%   with the alternative quantization by loop quantum gravity methods
%%   proposed by Thiemann in hep-th/0401172. This is done by
%%   expressing both in the 
%%   same operator algebraic formalism thereby also shading 
%%   some light on the Gupta-Bleuler procedure that leads to unitary
%%   operators implementing the diffeomorphism constraints in the Fock
%%   space construction. The main difference between the two approaches
%%   lies in the choice of expectation value functional which is more
%%   singular in Thiemann's case. By dropping the continuity assumption
%%   in the Stone-von-Neumann theorem, there an analogous choice can be
%%   made in the quantization of the harmonic oscillator and we discuss
%%   the (un)physical consequences of choosing he ``polymer'' state as
%%   done in loop quantum gravity.  
  We set up a unified framework to compare the quantization of the
  bosonic string in two approaches: One proposed by Thiemann, based on
  methods of loop quantum gravity, and the other using the usual Fock space
  quantization. Both yield a diffeomorphism invariant quantum theory.
  We discuss why there is no central charge in Thiemann's approach but
  a discontinuity characteristic for the loop approach to
  diffeomorphism invariant theories. Then we show the (un)physical
  consequences of this discontinuity in the example of the harmonic
  oscillators such as an unbounded energy spectrum. On the other hand,
  in the continuous Fock representation, the unitary
  operators for the diffeomorphisms have to be constructed using the
  method of Gupta and Bleuler representing the diffeomorphism group up
  to a phase given by the usual central charge.
\end{abstract}
%\submitto{\CQG}
\maketitle

\section{Introduction}
The most challenging problem in theoretical physics is to find a
consistent theory that encompasses General Relativity and quantum
physics. Currently, there are two major programs to attack this
challenge, string theory and loop quantum gravity. While string theory
is rooted in the high energy and particle physics tradition with focus
on scattering data and perturbative methods, loop quantum gravity's
emphasis is more on geometrical properties and diffeomorphism and
background independence due to its foundations in relativity.

With some notable exceptions, there has not been much interchange of
ideas between the two approaches which is partly due to the very
different mathematical languages used by the two camps. However
recently, there has been a conference at the Albert-Einstein-Institute
to bridge this cultural gap and as an effect of this meeting,
Thiemann\cite{Thiemann} has investigated the bosonic string world-sheet
theory using tools of canonical quantization typical of loop quantum
gravity. 

The world-sheet theory of a string seems to be the ideal toy model and
testing ground for approaches to quantum gravity: It is a proper field
theory with infinitely many degrees of freedom and it is (at least for
the classical theory) reparametrisation and therefore diffeomorphism
invariant. Thus it fits into the context of theories to which loop
quantum gravity methods apply. Yet, it allows for the ``conventional''
field theory treatment with Fock space operators. In the string theory
literature there are several equivalent formalisms (e.g. light-cone,
covariant or Gupta-Bleuler, BRST covariant, see for example
\cite{GSW}) that all lead to the same result: The quantized theory is
only consistent in the critical dimension, which is 26 for the bosonic
string. Otherwise, Lorentz- or conformal invariance are anomalous or
the BRST algebra does not close.

In his paper, Thiemann describes a quantization of bosonic closed
string theory that seems to work in any dimension. There are no anomalies
and thus no critical dimension. This is surprising from a string
theory point of view and it is the aim of this note to shed some light
on this result and what went into it.

Thiemann's treatment deviates in two ways from what is usually found
in string theory textbooks: First, there is a different
formalism. Instead of unbounded operators like $q$ and $p$ or
annihilation and creation operators, Weyl type
operators $e^{iq}$ and $e^{ip}$ are used because they are bounded and
thus continuous in the Hilbert space sense. Furthermore, this implies
that one deals with finite group transformations and not only
infinitesimal generators. In addition, care is taken to separate the
algebra of observables from its representations as operators
on a concrete Hilbert space. This is achieved by first constructing the
quantum algebra and then employing the GNS (Gelfand, Naimark, Segal)
construction. We should stress that so far the physical
content does not differ from the standard treatment except for
increased mathematical rigour than usually encountered in physicists'
dealings with functional analysis.

The second difference however is substantial: At some point in the
quantization procedure one has to choose an expectation value
functional or ``state''. While string
theorists usually choose (in most cases implicitly) the state
corresponding to the Fock-space vacuum, Thiemann chooses a different,
manifestly diffeomorphism invariant vacuum as it is typical for the
loop quantum gravity literature. In a similar context it is in fact
often argued that this choice is unique if one insists on
diffeomorphism invariance. However this so-called ``polymer'' state has some
physically  unfavourable properties: Technically speaking, the
representation is not (weakly) continuous and practically the state is
so singular that for example no momentum operators exist.

Indeed, the Fock-space vacuum as it is is not invariant under the
diffeomorphisms obtained from the quantization procedure. However,
what is technically required is not an invariant state but only
unitary operators on the Hilbert space that implement the action of
the diffeomorphisms on the algebra. In a second step (that has to be
taken in both approaches), one mods out the Hilbert space by the
action of the now unitarily implemented diffeomorphism group to obtain
the physical Hilbert space of invariant vectors. We will demonstrate
how to construct the unitary operators for the Fock representation in
this formalism. In effect, this is done by introducing normal ordering
and following the method of Gupta and Bleuler but the presentation will
differ significantly from the (often ad hoc) textbook treatment.

The price to pay for this construction of unitary implementers of the
diffeo\-mor\-ph\-isms is that they obey the group relations of the
diffeomorphism group only up to a phase. This is the manifestation of
the central charge in this formalism. Modding out these unitarities
including the phase would lead to an empty physical Hilbert space. To
circumvent this problem, we can however use the standard procedure: We
will take 26 copies of this theory with central charge one and add to
it the theory of a $bc$-ghost system with central charge $-26$ so that
in total we end up with a theory without anomalies. Thus, a consistent
string theory based on the (continuous) Fock space representation of
the quantum algebra has a critical dimension of 26.

In the end we will have learnt not very much about the usual string
theorists treatment of string quantization. We will have merely
reformulated well known results in a more precise formalism for
quantization maybe shedding some new lights on the inner workings of
the Gupta-Bleuler procedure; especially, we show that in the end, both
the positive and negative energy parts of the generators of symmetry
transformation are promoted to unitary operators. However, we can turn
Thiemann's argument around: This paper demonstrates that string theory
in the critical dimension in fact provides a highly non-trivial
example of a diffeomorphism invariant quantum theory that doesn't have
the unphysical properties of the quite singular ``polymer'' states
usually encountered in the loop quantum gravity literature.

To highlight the physical features of these polymer states we then
discuss the example of the one-dimensional harmonic oscillator: In
direct parallel to the case of the bosonic string (which in the end is
just an infinite collection of harmonic oscillators) there is the
standard Fock space quantization and an inequivalent one, based on the
singular polymer state. We find that in the polymer representation
only the ground state is stationary, all other states correspond to
scattering states. Furthermore, all other states have contributions of
arbitrarily positive and negative energies. This is in direct conflict
with the spectral properties of experimentally realised harmonic
oscillators showing that at least in the case of quantum mechanics the
polymer state is unphysical. As there are no quantum gravity or string
theory experiments, the question of which state for the quantum
algebra might be realised if at all in nature cannot be decided, but
these results for analogous quantum mechanical systems are quite
suggestive of what properties (such as weak continuity) one might
require in the quantization. These non-regular states however have an
application in the treatment of Bloch electrons, see for example
\cite{Strocchi}. In a periodic potential, the wave function has
to be periodic up to a phase. The total Hilbert space is then an
orthogonal sum over all possible phases and thus non-separable.

The structure of this paper is as follows: In the next section we
reformulate the quantization of the bosonic string in a language that
will allow us to compare the similarities and differences of the two
approaches. This includes a mathematical formulation of the Gupta
Bleuler construction of positive energy representations. Then follows
a chapter in which the same quantization procedure is applied to the
one dimensional case of the harmonic oscillator (the reader might
occasionally want to peek forward to this section as an illustration
of the formalism for the string). In this quantum mechanical example
we demonstrate that the two quantizations differ
observationally\footnote{See also the treatment in \cite{Willis} that
  comes to a different conclusion.} in their energy spectrum and thus
the Fock space quantization is clearly favoured experimentally.
In a final chapter we wrap up with some conclusions.

A final note to the mathematically cautious reader: Although we will
here probably employ a higher than usual standard of rigour, our
treatment will be purely algebraic. We will not discuss for example
convergence of sums that we write down (and most of the time suppress
necessary completions of infinite dimensional spaces) in order not to
burden the reader with too many notational details. However we are
positive that the missing details could be filled in without too much
work, for a treatment that contains these details we refer the reader
to \cite{Ottesen}.

\section{Canonical quantization of the string}
Here we would like to carry out the canonical quantization of the
closed bosonic string in all detail. 

Ideally, one would like to quantize the theory of unparametrised
strings, that is embeddings of $S^1\times \MR$ into target space.
However this leads to a much more involved theory (see for example
\cite{Rehren}) and we will instead quantize the theory of parametrised
strings, that is functions $X(x ,\tau)$ and then in the end
impose the constraint that physics is invariant of the parametrisation
we have chosen on the world-sheet. Thus we will deal with a gauge
system with unphysical degrees of freedom.

We will only be concerned with strings in flat target space. As in
this case all the dimensions of the target decouple we can for the
time being just treat an individual target coordinate and thus we do
not need any target space indices on $X$. Our convention for the size
of the closed string follows from
\begin{equation*}
X(x +2\pi,\tau)=X(x ,\tau).
\end{equation*}
We will take the Polyakov action
\begin{equation*}
  {\cal S}=\int\!\! d\tau\int_0^{2\pi}dx\,\left( -\partial_\tau X\partial_\tau
  X + \partial_x X\partial_x X\right)
\end{equation*}
as our starting point although this is not essential (\cite{Thiemann}
considers the Nambu-Goto action as well and ends up with the same
symplectic space). The canonical variables $X(x,0)$ and
$\partial_\tau X(x,0)$ have the Poisson-bracket
\begin{equation*}
  \{X(x,0),\partial_\tau X(x',0)\} = \delta(x-x').
\end{equation*}
As $X$ is periodic, all its information is also carried by the
currents
\begin{equation*}
  j^\pm(x) = \partial_\tau X(x,0)\pm \partial_x X(x,0).
\end{equation*}
Their Poisson-brackets
\begin{eqnarray*}
  \{j^+(x),j^+(x')\}=-\{j^-(x),j^-(x')\}=2\partial_x
  \delta(x-x'),\\
  \{j^+(x),j^-(x')\}=0
\end{eqnarray*}
decouple and therefore we can consider the $+$ and the $-$ components
independently. Thus from now we will drop the $\pm$ indices. Finally,
in order not to have to worry about distributions, we proceed from the
currents to their test-function duals, that is functions $f\colon
S^1\to\MR$ which we use to smear the currents:
\begin{equation*}
  f\mapsto j[f] = \frac1{\sqrt 2}\int_0^{2\pi}dx\, j(x) f(x).   
\end{equation*}
This mapping then induces the symplectic structure 
\begin{equation}
  \label{eq:PB}
  \sigma(f,g) = \big\{ j[f], j[g] \big\} = \int_0^{2\pi}dx f(x)\partial_x
  g(x)=\int fdg.
\end{equation}
Finally, by going to the rest-frame in target-space, we can assume $f$
to average to zero:
\begin{equation*}
  \int_0^{2\pi}dx f(x) = 0.
\end{equation*}
We wish then to quantize the space $M$ of real functions on
the circle with zero mean and the symplectic structure given by 
(\ref{eq:PB}). We will not need to specify a basis for this
space. However, most of the literature uses a language that
corresponds to a choice of basis. Often a Fourier decomposition that
is equivalent to a basis of the form
$(e^{inx})_{n\in\MZ^*}$ is used. Thiemann however chooses to take
characteristic functions of intervals in $S^1$ as his basis as these
bear some similarity with holonomy functionals in the loop approach to
quantum gravity \footnote{Note however that characteristic functions
  are discontinuous and the symplectic form (\ref{eq:PB}) is not
  well-defined on them.}. Of course, none of the results depends on the choice
of basis. 

Given any symplectic space $(M,\sigma)$, observables $F\colon M\to
\MR$ are in one-to-one correspondence to Hamiltonian vector fields
$X_F$ via
\begin{equation*}
  dF=\sigma(X_F,\cdot) \,.
\end{equation*}
Already at the classical level the vector fields do not commute, rather
they form a Lie algebra 
\begin{equation*}
  [X_F,X_G] = X_{\{F,G\}} \,.
\end{equation*}
To avoid complications with unbounded operators
later in the quantum theory, it convenient at this level not to deal
with infinitesimal symplectic transformations as given by Hamilton
vector fields but with their flows, i.e. finite symplectomorphisms
\begin{equation*}
  W(F) = \exp( \ad_{X_F}) = e^{\{F,\cdot\}} \,.
\end{equation*}
If $F$ and $G$ are canonical coordinates (that is, their Poisson-bracket
is a constant function, such as for $x$ and $p$) these maps can be
composed as
\begin{equation}
  \label{eq:cWeylalg}
  W(F)\circ W(G)=e^{\frac 12\{F,G\}} W(F+G)
\end{equation}
with the help of the Campbell-Baker-Hausdorff formula. This algebra is
easily recognised as the algebra obeyed by Weyl operators. Indeed,
quantization of our classical system is the promotion of this
classical algebra to a quantum algebra. With our $f\colon S^1\to\MR$
as above, the quantum algebra $\cal W$ is generated by elements $W(f)$ with
relations  
\begin{equation}
  \label{eq:Weylalg}
  W(f)W(g)=e^{\frac i2\sigma(f,g)} W(f+g)
\end{equation}
and conjugation $W(f)^*=W(-f)$ and can be checked to be a C*-algebra,
the algebra of canonical commutation relations. There is a unique norm
on this algebra but we do not use it explicitly in this paper. 

It is important to realise that although the classical symplectic maps
$W(F)$ exist for all functions on phase space, at first we promote
only the elementary ones (i.e. linear functionals on $M$; 
in the case of mechanics, these are $x$ and $p$ but not powers of
them) to quantum operators that obey the Weyl algebra 
(\ref{eq:Weylalg}). 

Noether's theorem says that classically, all one-parameter
groups of symplectomorphisms are inner, i.e. they can be written as
$W(F)$ where $F$ is the related conserved charge. However this does
not automatically lead to an inner transformation in the quantum
theory. This is because there is not necessarily an operator that
represents $F$ on the Hilbert space as in general $F$ is not one of the
elementary observables for which we defined quantum operators in the
beginning. In fact, the rest of this section is concerned with
constructing such operators for two Hilbert space realisations of the
quantum algebra (\ref{eq:Weylalg}).

As in Lie algebra theory where one first studies the abstract algebras
and then in a second step their representations in order to separate
properties of the algebra from properties of specific representations,
the same can be done in quantum theory: First we analyse the abstract
C*-algebra of observables and only then study its representations in
terms of linear operators on a Hilbert space. This split is usually
not considered necessary for quantum mechanics (and therefore omitted
in most textbooks) as the Stone-von Neumann theorem guaranties that
for the Weyl algebra of a finite number of degrees of freedom the
usual Schr\"odinger representation is the only one possible. We will
however later have to reconsider this short-cut as one might want to
drop one of the assumptions of this theorem (see section 3).  The
situation is however different in quantum field theory where the
C*-algebras involved admit inequivalent representations, giving rise
to the theory of superselection sectors.

In the case of the bosonic string, again at the level of abstract
C*-algebras, the usual textbook treatment and the approach of
\cite{Thiemann} agree, it is only at the level of representations where they
differ. As we will see later, this difference could already be made
for quantum mechanical systems such as the one-dimensional harmonic
oscillator. 

It is a classic result by Gelfand, Naimark, and Segal (for an extended
discussion see for example \cite{Haag}) that representations of
C*-algebras $\cal A$ arise from states via the construction named
after them: A state is a linear functional $\omega:{\cal A}\to\MC$
that should be thought of as assigning an expectation
value to each observable. One requires that, for an algebra with a unit
$\ID\in{\cal A}$, the state is 
normalised to $\omega(\ID)=1$, and that it is positive for positive
elements of the algebra:
\begin{equation*}
  \forall A\in{\cal A}\colon\quad \omega(A^*A)\ge 0.
\end{equation*}

Given such a state, we can define ${\cal J} = \{A\in{\cal
  A}|\omega(A^*A)=0\}$ (which in the cases we will be interested in
can be checked to be trivial, ${\cal J}=\{0\}$) and verify with the help of the
Cauchy-Schwarz inequality that it is an ideal. Then we can define a
  vector space ${\cal H}={\cal A}/{\cal J}$. This carries a representation
\begin{equation*}
  \rho(A)|B\rangle=|AB\rangle
\end{equation*}
where we use $|B\rangle$ to indicate the class of $B$ in ${\cal
  A}/{\cal J}$. To make $\cal H$ into a Hilbert space we also need a
scalar product and this is given by 
\begin{equation*}
  \langle A|B\rangle = \omega(A^*B).
\end{equation*}

This however, is not the whole story, as the the algebra of the
$W(f)$ is that of parametrised strings but we want our final quantum
theory to be invariant under reparametrisations. So let $S:S^1\to S^1$
be a reparametrisation of the circle. It can be pulled back to the $f$
via
\begin{equation*}
  (Sf)(x) = f(S(x))
\end{equation*}
and furthermore to the Weyl operators
\begin{equation*}
  S:W(f)\mapsto \alpha_S(W(f)) = W(Sf). 
\end{equation*}
From the second form of the symplectic structure (\ref{eq:PB}) it
follows that $S$ induces a sym\-plec\-to\-mor\-phism $\sigma(Sf,Sg) =
\sigma(f,g)$. As it respects the Weyl relation (\ref{eq:Weylalg}),
$\alpha_S$ is an automorphism of the quantum algebra. Moreover,
diffeomorphism of the circle can be composed and form a group. Obviously, the
map $\alpha\colon Diff(S^1)\to Aut({\cal W})$ that maps $S$ to
$\alpha_S$ is a group homomorphism:
\begin{equation*}
  \alpha_{S_1}\circ\alpha_{S_2}=\alpha_{S_1\circ S_2}.
\end{equation*}
The crucial task
will turn out to be to give a map from the group of C*-algebra
automorphisms $\alpha_S$ to the group of unitary operators $U(S)$ on the
representation Hilbert space $\cal H$, such that 
\begin{eqnarray*}
  U(S)\rho(A)U(S)^{-1}=\rho(\alpha_S(A))
\end{eqnarray*}
and then mod out by these
unitary transformations to obtain the physical Hilbert space.
Ideally, one would want the map $U:\alpha_S \to U(S)$ to be a group
homomorphism, so that $U(S) U(S') = U(S \circ S')$, 
but we will see that this property in general can not be satisfied:
This will be
the place where the central charge appears in this description. But we
should note that the central charge is really a property of the
representation and not of the quantum algebra!

If in general one has a C*-algebra ${\cal A}$ and an automorphism
$\alpha$ acting on it, there is no canonical way to promote it to a
unitary operator on the representations. However, if the state
$\omega$ that GNS-constructs this representation is invariant
($\omega(\alpha(A))=\omega(A)$ for all $A$), then there is a unitary
right at hand: We can define $U$ to act as
$U|A\rangle=|\alpha(A)\rangle$. This is unitary as invariance of the
state induces invariance of the scalar product:
\begin{eqnarray*}
  (U\langle A|)(U|B\rangle) &= \langle\alpha(A)|\alpha(B)\rangle 
  =\omega(\alpha(A)^*\alpha(B)) \\
  &= \omega(\alpha(A^* B)) 
  = \omega(A^*B) 
  = \langle A|B\rangle.
\end{eqnarray*}
As $U$ arises from a pull-back of $S$, it preserves the group
structure of the sym\-plec\-to\-mor\-phisms $S$, that is it is a group
homomorphism from the group of $S$'s to the unitary group of $\cal H$
and thus unitarities constructed this way do not give rise to central
charges. However, if the state is not invariant, this construction of
unitary implementers is not available and the group property is not
automatic.

So far, we have only discussed the abstract algebra and have not yet
decided to proceed to the representation theory, that is we have not
decided for a (``vacuum'') state or, equivalently, a Hilbert space on
which the algebra acts. Up to this point, the only difference between
the standard treatment and Thiemann's is in the language used not in
the content. However, the two approaches differ in their choice of
state. Thiemann makes this explicit in his paper whereas this choice
is only implicit in the usual textbook treatments.

Thiemann chooses a state of ``polymer type'' similar to the state that
is used in LQG, specifically, he takes
\begin{equation*}
  \omega_P(W(f)) = 
  \cases{1&if $f=0$\cr 0&else.}
\end{equation*}
This choice has the obvious advantage of being invariant under
diffeomorphisms of the circle as a reparametrisation maps
non-zeros functions to non-zero functions. Thus one can apply the
above construction for the unitary implementers on the Hilbert
space. In fact, in the context of gravity this appears to be the
unique diffeomorphism invariant state\cite{Sahlmann,
  Thiemann}. However, for an invariant theory, we only need the
unitary operators (and their kernels) and  an invariant state provides
those canonically but as we will see below, in other states they can
be defined as well.

On the other side, this state has one unusual property: It is not
continuous in the argument $f$! This implies that we cannot take the
derivative of the Weyl operators $\rho(W(f))$ to obtain operators for
the field (``position operator'' in mechanics) or its momentum. The
action of the Weyl operators on this Hilbert space is so singular,
that the field and the field-momentum cannot be defined. Furthermore,
the Hilbert space constructed from $\omega_P$ is huge in the sense
that it is not separable. Thus it does not have a (countable)
orthonormal system. Rather, its basis is labelled by (the continuum)
of functions $f$ and the overlap $\langle W(f)|W(g)\rangle$ vanishes
unless $f=g$. 

Even in quantum mechanics, (weak) continuity is an assumption of the
Stone-von Neumann theorem that guarantees the choice of position
and momentum operators to be unique (up to unitary equivalence such as
change from position to momentum representation). If that assumption
is dropped there are non-standard quantizations of ``polymer''-type
with non-separable Hilbert-spaces as well and we will study the
physical consequences of such a quantization of the harmonic
oscillator in section 3.

It should be said however, that this non-separable Hilbert space is
only an intermediary. In \cite{Rovelli,Zapata} it is showed that once the
constraints have been modded out (at least in the case of 3+1 gravity)
the physical Hilbert space is again separable.

The usual Fock representation in contrast comes from a state that is
continuous. To define it we need some more input. Namely, as the Fock
representation is a positive energy representation (and negative
energy modes of all the fields annihilate the vacuum) we have to
introduce a way to distinguish positive and negative energy modes. At
this point, most textbooks now introduce the Fock vacuum (that is
annihilated by negative modes) in an ad hoc way and then later proceed
to impose (under the names of Gupta and Bleuler) only the positive
energy half of the constraints for ``quantum consistency''. Here, we
will spend some more time on the details of this procedure and show
how it fits into the general framework of deformation quantization and
GNS-construction. Our treatment follows along the lines of
\cite{Ottesen}.

The distinction between positive and negative modes can be encoded in
the definition of a complex structure $J$ for the functions $f$ that
turns the symplectic space into the complex one-particle Hilbert
space.  Specifically, we require $J$ to square to $-\ID$
and to be skew with respect to the symplectic structure:
\begin{equation*}
  \sigma(Jf,g)=-\sigma(f,Jg)
\end{equation*}
for all $f$ and $g$. In addition we require
\begin{equation*}
  \sigma(f,Jf)\ge0
\end{equation*}
for all $f$. The conventional choice amounts to 
\begin{equation*}
  (Jf)(x) = \frac1{2\pi}\int_0^{2\pi} dy \, f(y) \cot \frac
  12 (y-x) 
\end{equation*}
 where the integral is evaluated using the principal value
 prescription. This form becomes more familiar in a Fourier
 basis. Defining
 \begin{equation}
   \label{eq:modes}
   f_n=\frac1{2\pi} \int_0^{2\pi} dx\, f(x)e^{-inx}
 \end{equation}
the complex structure acts as
\begin{equation*}
  J\colon f_n\mapsto \sgn(n) i f_n.
\end{equation*}
So the positive energy modes are in the eigenspace of $J$ with
eigenvalue $+i$. Now, we can complexify the space of $f$'s and with
the scalar product
\begin{equation}
  \langle f|g\rangle = \sigma(f,Jg) + i \sigma(f,g)
\end{equation}
(which is sesquilinear because of the properties of $J$) it becomes
the one particle Hilbert space. After all these preparations, second
quantization amounts simply to give the state that corresponds to the
Fock vacuum:
\begin{equation}
  \label{eq:Fockstate}
  \omega_F(W(f)) = e^{-\frac14 \langle f|f\rangle}.
\end{equation}
Obviously, this choice is continuous in $f$ and, using the CBH
formula, the reader can easily convince herself for example in the
Fourier basis that is just the standard vacuum expectation value of
the operator $\exp(\sum_n f_n a_n)$. As the Fock state is weakly
continuous, we also take derivatives of the Weyl operators and define
the usual hermitian field operators as
\begin{equation}
  \label{eq:pidef}
  \pi(f) = \left.\frac d{d\lambda}\right|_{\lambda=0} \rho(W(\lambda f))
\end{equation}
and creation and annihilation operators as 
\begin{equation}
  \label{eq:adef}
  a^*(f)=\frac1{\sqrt 2}(\pi(f)-i\pi(Jf)),\qquad a(f)=\frac1{\sqrt
  2}(\pi(f)+i\pi(Jf)).
\end{equation}
If we write the vacuum vector $|W(0)\rangle=|\ID\rangle$ in the usual way as
$|0\rangle$, they act in the usual way on vectors of the form
\begin{equation*}
  a^*(f_1) \vee\cdots \vee a^*(f_n)|0\rangle.
\end{equation*}
($\vee$ indicates symmetrisation) with commutation relations
\begin{equation*}
  [a(f),a^*(g)]=\langle f|g\rangle \cdot 1.
\end{equation*}
as follows from (\ref{eq:Weylalg}), (\ref{eq:pidef}) and
(\ref{eq:adef}).

Now that we have defined the Fock state and understand the structure
of the corresponding Hilbert space we can investigate the action of
diffeomorphisms. Given a diffeomorphism $S$ acting on $f$, we would
like to define a corresponding unitary operator $U(S)$ on the Fock
space. As with the Fock space derivatives exist, we can as well work
with the infinitesimal version by writing $S=e^A$. As a
diffeomorphism $S$ leaves the symplectic structure invariant, it follows
that $A$ is skew ($\sigma(Af,g)=-\sigma(f,Ag)$). From (\ref{eq:Fockstate}),
it follows that if $A$ anti-commutes with $J$ the norm $\langle
f|f\rangle = \sigma(f,Jf)$ is annihilated by $A$ and thus the state
$\omega_F$ is invariant under $S$. 

Unfortunately, this is not always the case. $S$ in general leads
to a Bogoliubov transformation and those do not leave the vacuum
invariant. In the general situation, we can split $A=A_1+A_2$ as
\begin{equation*}
  A_{1\atop 2}= \frac 12 (A\mp JAJ)
\end{equation*}
such that $A_1$ anti-commutes and $A_2$ commutes with $J$. Physically,
this means that $A_1$ maps positive to positive and negative to
negative energy modes whereas $A_2$ mixes the two. As we saw, $A_1$
leaves $\omega_F$ invariant so we can find a unitary $U(e^{A_1})$
(and its derivative $dU(A_1)$ in the canonical way. Concretely, it
acts as
\begin{equation*}
  U(e^{A_1})\left(  a^*(f_1) \vee\cdots \vee a^*(f_n)|0\rangle\right)
  =   a^*(e^{A_1}f_1) \vee\cdots \vee a^*(e^{A_1}f_n)|0\rangle.
\end{equation*}
The complex anti-linear part $A_2$ also has a uniquely
defined\cite{Shale} action in the Fock space, which is, however,
slightly more involved. It turns out \cite{Ottesen} that in our case
$A_2$ is a self-adjoint Hilbert-Schmidt operator. Thus there exist
orthogonal sets $\{u_i\}_{i\in \MN }$ and $\{v_i\}_{i\in \MN }$ that span
the range of $A_2$ such that the action of $A_2$ can be written as
\begin{equation}
  \label{eq:HS}
  A_2f=\sum_{i\in \MN } \langle f|v_i\rangle u_i=\sum_{i\in \MN } \langle
  f|u_i\rangle v_i 
\end{equation}
where the second equation follows from self-adjointness. This
representation leads us to the definition of a creation operator
$a^*(A_2)$ acting as
\begin{eqnarray*}
  a^*(A_2)&\left(a^*(f_1) \vee\cdots \vee a^*(f_n)|0\rangle\right)
  =\\
  &\sum_{i\in \MN } a^*(u_i)\vee a^*(v_i)\vee a^*(f_1) \vee\cdots \vee
  a^*(f_n)|0\rangle 
\end{eqnarray*}
and an annihilation operator as its adjoint
$a(A_2)=(a^*(A_2))^*$. Altogether, we obtain an anti-hermitian
operator on the Fock space by
\begin{equation*}
  dU(A_2)=\frac 12(a(A_2)-a^*(A_2))
\end{equation*}
and finally $dU(A)=dU(A_1)+dU(A_2)$, that can be exponentiated to
yield the unitary operator
\begin{equation*}
  U(e^A)=\exp(dU(A)) \,.
\end{equation*}
From these definitions it follows that
\begin{equation*}
  dU(A)|0\rangle= -\frac 12 a^*(A_2)|0\rangle
\end{equation*}
as the $A_1$-part leaves the vacuum invariant and the annihilation
part of $dU(A_2)$ vanishes on the vacuum. Thus we have $\langle
0|dU(A)|0\rangle =0$ and 
\begin{eqnarray*}
  \langle 0|dU(A)dU(B)|0\rangle &= -\frac 14\langle0|a(A_2)a^*(B_2)|0\rangle \\
  &=-\frac 12 \Tr(B_2A_2)
\end{eqnarray*}
where the trace exists because the anti-linear parts $A_2$ and $B_2$
are Hilbert-Schmidt and the last equality follows from the spectral
form (\ref{eq:HS}). Finally, it is easy to check that $dU(A)$ actually
implements $A$ on the Fock space, i. e.
\begin{equation*}
  [dU(A),\pi(f)]=\pi(Af)
\end{equation*}
which is the infinitesimal form of
\begin{equation*}
  U(S)^{-1}W(f)U(S)=W(S^{-1}f).
\end{equation*}
Obviously, this equation would specify $U(S)$ only up to a phase and
thus the uniqueness of the Weyl algebra implies that the group
relations from compositions of diffeomorphisms hold in the Fock space
up to a phase. Infinitesimally, this means
\begin{equation}
  [dU(A),dU(B)]=dU([A,B]) + \phi(A,B)\ID
\end{equation}
where $\phi$ is purely imaginary. Taking the vacuum expectation value
of this equation then gives us
\begin{equation}
  \phi(A,B) = \frac 12\Tr([A_2,B_2]).
\end{equation}
For the trace of a commutator not to vanish it is necessary that the
one particle Hilbert space in which $A_2$ and $B_2$ act is infinite
dimensional. So we find that in the Fock representation of the Weyl
algebra $\phi$ is an obstruction to the unitary implementation of the
classical diffeomorphism group. 

Let us compute this obstruction in the concrete case of the
closed string. Diffeomorphisms are generated by vector fields
\begin{equation*}
  L_k = e^{ikx}\frac d{dx}
\end{equation*}
where for simplicity we complexify from the very beginning. They act
on the Fourier components (\ref{eq:modes}) as $L_k f_n=
-nf_{n+k}$. In this basis, the scalar product is given by
\begin{equation*}
  \langle f|g\rangle = \sum_{n>0} n\bar{f_n}g_n.
\end{equation*}
First we compute
\begin{eqnarray*}
  (L_k)_2f_n&=\frac 12 (L_kf_n+JL_kJf_n)\\
  &= \frac 12(-nf_{n+k}+i \sgn(n)JL_kf_n)\\
  &= \frac n2 (\sgn(n)\sgn(n+k)-1)f_{n+k},
\end{eqnarray*}
finding it non-zero only if $n$ and $n+k$ have different
signs. Setting $d(n,k)= \sgn(n)\sgn(k)-1$ we finally compute
\begin{eqnarray*}
  \Tr\left([(L_n)_2,(L_k)_2]\right) &=\sum_{j>0}\frac 1j\langle
  f_j|[(L_n)_2,(L_k)_2]|f_j\rangle \\
&=\sum_{j>0}\langle f_j|\left( -(L_n)_2d(k,j)|f_{j+k}\rangle+(L_k)_2
  d(n,j) |f_{j+n}\rangle \right)\\
&=\sum_{j>0}j \delta_{k,-n}\big((j+k) d(k,j) d(-k,k+j)\\
&\qquad - (j-k) d(-k,j)
  d(k,-k+j)\big). 
\end{eqnarray*}
This sum has only $|k|-2$ non-vanishing terms and yields
\begin{equation*}
  \phi(L_n,L_k) = \frac 1{12} n(n^2-1)\delta_{k,-n} 
\end{equation*}
which reflects the well known fact that the free boson has central
charge 1. Note that $L_n$ and $L_{-n}$ are adjoint to each other so
that the anti-hermitian combinations $\frac 12(L_n - L_{-n})$ and
$\frac i2(L_n+L_{-n})$ that are exponentiated to give the unitary
operators actually have a purely imaginary central component in their
commutator. Note that our calculation was finite at all stages and did
not require any regularisation.

The Hilbert spaces we have constructed so far are by themselves not
physical: The diffeomorphisms of the string do not act trivially on
them and they would describe the quantum theory of parametrised
strings. What we have done in this section is to construct unitary
operators on these kinematical Hilbert spaces that implement the
action of the diffeomorphisms. In a final step, these have to be
divided out and only the vectors that are left invariant by the $U(S)$
are the physical states. 

The appearance of the central charge in the Fock space implying that
multiples of the unit operator are in the symmetry algebra is deadly
for this second step: Invariance would require that one of the physical
state conditions would be
\begin{equation*}
  e^{\phi}|\psi\rangle = |\psi\rangle
\end{equation*}
where the central $\phi$ is an imaginary number. This condition
would imply that there are no physical states. The way out of this
problem however is well known: The quantization of only one $X$ is not
consistent, one has to add further fields such that the total central
charge vanishes. In the case of the bosonic string this is done by
taking 26 copies of the theory we have studied here and add a
similarly quantized $bc$ ghost system that provides a central charge
of $-26$ so the total sum of the symmetry implementers $L_n$ do not
have a central charge and thus no multiples of the unit operator in their
algebra.

In this section we have discussed the two inequivalent Hilbert space
representation of the quantum C*-algebra of the bosonic string,
arising from the polymer state as suggested by Thiemann and the Fock
state that underlies the usually treatment of the theory. We mentioned
that Thiemann's $\omega_p$ is not (weakly) continuous and therefore
many of the usual operators do not exist on the polymer Hilbert
space. In the next section we will discuss the physical consequences
of this fact in the simpler (and of course experimentally accessible)
but completely analogous case of the harmonic oscillator. 

We have made no attempt here to investigate whether there are further
inequivalent states on the algebra of the $W(f)$. It is however well
known\cite{Ottesen,Buchholz} that if one insists on continuity there is a
theorem similar to the one by Stone and von Neumann in the quantum
mechanical case, that guarantees the uniqueness of the Fock
representation. So even the apparent choice of complex structure $J$
we have had to make above has no room for alternatives that lead to
inequivalent Hilbert spaces.

\section{The harmonic oscillator as a testing ground}
It has been argued\cite{coffeetable}, that as nobody has so far
observed a quantum string in an experiment, there is no empirical data
on which to base the decision for one or the other state leading to
different physical properties of the quantized string. Especially, the
fact that the diffeomorphism group acts on the polymer Hilbert space
without a central charge (as the state $\omega_P$ on which it is based
is invariant under diffeomorphisms) seems to make this quantization
far more generally applicable. In this paper, we have only discussed a
single $X$ and therefore target space symmetries (especially
Poincar\'e invariance) did not play a r\^ole here, but Thiemann
describes in his paper that as he lets these symmetries act trivially
on the internal theory of the string, the oscillations of the string
carry no momentum in his theory and there are strings of all
rest-masses like it is the case for point-particles. In particular, he
can find a tachyon-free bosonic string. We will
not comment here on this property of Thiemann's string but just say
that all these features seem to make his model far more attractive
as a physical theory than the ordinary string.

Again, up to the foreseeable future, there are no experiments in sight
that directly test fundamental string theory or quantum gravity so one
might think there are no empirical preferences for the choices one has
to make during the quantization procedure. However, we will argue in
this section that one has exactly the same choices in the quantization
of ordinary quantum mechanical systems like the harmonic oscillator or
the hydrogen atom that are experimentally tested on a daily basis and
that there the choice for the polymer state leads to unphysical
consequences.

Before we start, let us however warn the reader that there is one
significant difference between the string and quantum gravity on one
side and the mechanical systems on the other: The later are not
gauge systems with redundant degrees of freedom that have to be modded
out to turn the kinematical Hilbert space into the physical one. We
make no attempt here to understand the structure of the physical
Hilbert space (in the Fock space case it is known to be generated by
DDF states for a recent development, see \cite{Schreiber}) and
physical criteria should strictly only be applied to what is left
after the gauge freedom has been removed but we still think it is
useful to illustrate  the
physical consequences of non-continuous representations in the case of
the harmonic oscillator. 

In the case of gravity, it has been argued that polymer-type states
are the only states on the quantum algebra of observables that are
invariant under diffeomorphisms. But it is one of the main points of
this paper to emphasise that invariant states are not necessarily
needed and that at least in the case of the bosonic string there are
other ways to construct unitary implementers of the symmetry group
that act on a much less singular Hilbert space than the one obtained
from $\omega_P$. 

Our quantization procedure for one-dimensional quantum mechanics and
the harmonic oscillator specifically will be very similar to the
discussion of the bosonic string in the previous
section. Notationally, it just consists in the replacement $f\mapsto
z$. The only difference is that now our symplectic space has finite
dimension. Concretely, we take it to be $\MC=\MR^2$, and combine
position and momentum into real and imaginary parts of a complex
number $z$. The usual symplectic form is
\begin{equation*}
  \sigma(z,z')= \Im(z\bar{z'}).
\end{equation*}
As before, instead of using position and momentum directly, we
exponentiate them to $W(z)=\exp(iz)$ to obtain bounded (unitary) Weyl
operators after quantization. So, our quantum algebra is generated by
linear combinations of operators $W(z)$ and there is the canonical
``commutation'' relation
\begin{equation*}
  W(z_1)W(z_2) = e^{\frac i2\sigma(z_1,z_2)}W(z_1+z_2).
\end{equation*}
The formal similarity to the bosonic string should not come as a
surprise as the latter is a free theory that is formally the sum of
infinitely many harmonic oscillators. Hilbert spaces on which this
C*-algebra is represented are again obtained from states (expectation value
functionals) with the help of the GNS-construction. The fact that
usually one does not make a difference between the elements of the
abstract algebra and operators on a Hilbert space is justified by the
Stone-von Neumann theorem that states that there is only one
representation of the Weyl algebra that is continuous in $z$. It
is based on the Fock vacuum
\begin{equation*}
  \omega_F(W(z)) = e^{-|z|^2}.
\end{equation*}
The Hilbert space is then described by acting on the vacuum vector
$|0\rangle$ associated to $W(0)=\ID$ by operators $\rho(W(z))$ (and
taking the completion with respect to the norm coming from the scalar
product $\langle z_1|z_2\rangle = \omega_F(W(z_1)^*W(z_2))$). This
Hilbert space is the usual $L_2(\MR)$ on which for real $x$ the
operator $\rho(W(x))$ translates functions by $x$ and the operator
$\rho(W(ix))$ multiplies functions by $\exp(ix)$.

Position and momentum operators combined into hermitian and
anti-hermitian parts of an operator are then derivatives of Weyl
operators
\begin{equation*}
  \pi(z) = \left.\frac d{d\lambda}\right|_{\lambda=0} \rho(W(\lambda z))
\end{equation*}
and creation and annihilation operators 
\begin{equation*}
  a^*(z)=\frac 1{\sqrt 2} (\pi(z)-i\pi(iz))\qquad a(z)=\frac 12 (\pi(z)+i\pi(iz)).
\end{equation*}
It is easy to check that in fact $a(z)$ annihilates the vacuum
$|0\rangle$. Thus the states $|z\rangle=\rho(W(z))|0\rangle$ are the coherent
states
\begin{equation*}
  |z\rangle = e^{-\frac 12|z|^2} e^{za^*}|0\rangle.
\end{equation*}
So far, we described general one dimensional quantum mechanics without
reference to a specific system. Classically the dynamics is specified by
a Hamilton function, $H=\frac 12(p^2+x^2)$. As before, we will however
proceed to the integrated flow in phase space rather than the infinitesimal
generator (as that might not exist in the quantum theory or at least
be an unbounded operator). The time evolution of the harmonic
oscillator is just rotation in phase space:
\begin{equation*}
  U(t)\colon z\mapsto e^{it}z.
\end{equation*}
The vacuum state $\omega_F$ is invariant under the corresponding
automorphism $\alpha_t(W(z))=W(e^{it}z)$ of the Weyl
algebra. Therefore we directly obtain the unitary implementers on the
Hilbert space:
\begin{equation*}
  U(t)|z\rangle=|e^{it}z\rangle
\end{equation*}
Using the chain rule, it follows that the eigenstates of this time
evolution are just the usual ones:
\begin{equation*}
  U(t)(a^*(z))^N|0\rangle = e^{iNt}(a^*(z))^N|0\rangle
\end{equation*}
This concludes our discussion of the standard Schr\"odinger
quantization of the harmonic oscillator. Next we want to contrast
these properties with a quantization that is based on a state that is of
the same ``polymer'' form that Thiemann used in his quantization of
the bosonic string and that parallels states used in the loop
quantization program of gravity. We define
\begin{equation*}
  \omega_P(W(z))=\cases{1&if $z=0$\cr 0&else.}
\end{equation*}
This is a slightly different choice than the one used in \cite{Willis}
but ours has the advantage of being invariant under the time evolution of
the harmonic oscillator, thus making at least the ground state
stationary. Again, this state is not continuous in $z$ and thus the
derivatives needed to define position and momentum operators $\pi(z)$
and thus creation and annihilation operators $a^{(*)}(z)$ do not
exist. 

This choice of state leads to a rather unusual Hilbert space: The
states $|z\rangle = W(z)|0\rangle$ are all orthogonal as long as their
arguments $z$ differ. After taking the completion with respect to the
GNS norm, elements of the Hilbert space are functions on $\MC$ that
are non-zero at at most countably many points $(z_i)_{i\in\MN}$ 
\begin{equation}
\label{state}
  \phi=\sum_{i\in\MN}\phi_i |z_i\rangle
\end{equation}
and
which are $l_2$-normalisable:
\begin{equation*}
  \sum_{i\in\MN}|\phi(z_i)|^2=\sum_{i\in\MN}|\phi_i|^2<\infty.
\end{equation*}
The scalar product of two such functions $\psi$ and $\phi$ is a sum
over the points on which both of them are non-zero
\begin{equation*}
  \langle\psi |\phi\rangle=\sum_{i\in\MN}\overline{\psi(z_i)}\phi(z_i).
\end{equation*}
It is interesting to note, that this Hilbert space (including its
scalar product) does not contain any information about the topology of
$\MC$ anymore: Any bijective (possibly discontinous) map $\MC\to\MC$
that fixes the origin leaves the state $\omega_P$ invariant
and leads to unitarily equivalent Hilbert space, that is the Hilbert
spaces are ``the same''.

It is clear that translation operators $\rho(W(z))$
acting on such functions do not have well defined derivatives and thus
position and momentum operators cannot be defined. Furthermore, as
the $|z\rangle$ are orthogonal for all $z\in\MC$, this Hilbert space
is not separable but we will not dwell on this point as in the case of
gravity the physical Hilbert space will again be separable after the
constraints have been modded out \cite{Rovelli,Zapata}.

Far more important are the physical properties of the harmonic
oscillator in this representation: Again, the polymer state is
invariant under the time evolution $\alpha_t(W(z))=W(e^{it}z)$,
therefore there are canonically given unitary operators
$U(t)$. However, as again they do not have a derivative with respect
to $t$, there is no Hamilton operator and we cannot directly discuss
the energy spectrum.

In \cite{Willis}, this problem is circumvented by introducing an ad
hoc scale and defining a Hamilton operator in terms of finite
difference operators at that scale instead of derivatives. However,
this ad hoc procedure does not connect with the classical time
evolution of the harmonic oscillator. In this paper, we take the point
of view that the time evolution operators 
\begin{equation*}
  U(t)|z\rangle = |e^{it}z\rangle
\end{equation*}
follow from the correspondence principle and we have to investigate
their properties (this is in the spirit of deformation quantization,
for an introduction, see for example \cite{Hirshfeld}).

First of all, the only vector in the Hilbert space that describes a
stationary state and transforms with a phase under this time evolution
is $|0\rangle$. Thus it is fair to say that $|0\rangle$ is the only
bound state of this version of the harmonic oscillator. All other
vectors describe ``scattering'' states.

The closest to what would be a Hamiltonian is  the hermitian operator
\begin{equation*}
H_\epsilon = \frac{U(\epsilon)-U(-\epsilon)}{2i\epsilon }  
\end{equation*}
which would converge to the Hamiltonian if the limit $\epsilon\to 0$
existed. For fixed $\epsilon$, $H_\epsilon$ can have eigenstates.
However they will not be eigenstates for all $U(t)$ as it is the case
in the Schr\"odinger representation. Eigenstates of $H_\epsilon$ are
thus not stationary, they are just periodic up to a phase with period
$2\epsilon$.

For generic, non-zero $\epsilon$ the expectation value of $H_\epsilon$
is zero in any state. However, when acting on a normalised
state $\phi$ as in (\ref{state}) it produces a state of norm
\begin{equation*}
  \| H_\epsilon\phi\|^2=\frac 1{2\epsilon^2}\sum_{i\in\MN, z_i\ne
  0}\bar\phi_i\phi_i. 
\end{equation*}
Thus we find that the expectation  value of $H_\epsilon^2$ diverges as the
limit $\epsilon\to 0$ is attempted: All states except for the ground
state have diverging energy and also the energy spectrum is not
bounded from below. This is clearly in conflict with the energy
spectra of harmonic oscillators found in nature \footnote{This result
  is not surprising in view of the fact that the polymer state can be
  seen as a thermal state in the limit of infinite temperature, see
  \cite{Strocchi}.}.   
Therefore we have to
conclude that at least for the harmonic oscillator, the quantization
based on the polymer state is empirically not correct. The polymer
state is too singular and thus gives an unphysical Hilbert space for
quantum mechanics. 

Of course, it would be too quick to extrapolate this result for the
harmonic oscillator directly to the bosonic string or even to the
quantization of gravity but it shows that one should have good reasons
to depart from the quantization scheme that requires continuity of the
states that was successful in the experimentally tested case of
quantum mechanics.

\section{Conclusions}
In this note, we have put both the usual Fock space quantization of
the bosonic string and Thiemann's alternative approach into a common
formalism that exposes at which points the two treatments are
equivalent and at which points they differ. As long as the algebra of
observables is considered as an abstract C*-algebra, both approaches
are completely parallel but they differ in the choice of a
representation of this algebra as operators on a Hilbert
space. 

Thiemann chooses a representation based on a polymer state as it is
always done in the loop approach to quantum gravity. This state has
the advantage that it is invariant under diffeomorphisms and therefore
directly leads to unitary operators in the Hilbert space representing
the diffeomorphism group. However, this representation is not
continuous. Therefore infinitesimal generators like field operators
and field momenta cannot be defined as derivatives of Weyl
operators. Furthermore, the kinematic Hilbert space constructed this
way is not separable.

The usual representation is constructed as a Fock space which has the
continuity property that the polymer representation lacks. Therefore
the usual creation and annihilation operators can be defined. On the
other hand, the Fock vacuum is not invariant under diffeomorphisms as
they act as Bogoliubov transformations. Nevertheless, unitary
operators for the diffeomorphisms can be defined by a variant of the
Gupta Bleuler procedure. Generically, this leads to a representation of
the diffeomorphism group up to a phase which is the integrated form of
a central charge. However, in the critical dimension this vanishes and
a diffeomorphism invariant physical Hilbert space can be obtained. 

One could argue on general grounds, that because of Fell's
theorem\cite{Haag} the choice of a representation is immaterial, as it
cannot be determined by any finite number of measurements of finite
precision. However, for any two different states one can always find an
experiment that distinguishes between them, so the two representations
should be regarded as physically inequivalent. 

In a later chapter, we discussed that if one drops the requirement of
weak continuity and thus circumvents the theorem by Stone and von
Neumann, exactly the same choice of representations exists in quantum
mechanics and leads to physically inequivalent quantizations of the
harmonic oscillator. Especially the energy spectrum in the polymer
representation differs significantly from the usual experimentally
observed spectrum.

As mentioned in the introduction, this discussion is not only relevant
to string theory. The world-sheet theory of the bosonic string appears
to be a simple but non-trivial testing ground for the quantization of
diffeomorphism invariant theories. As such, the Fock space treatment
of string theory text books can be interpreted as providing a
canonical quantization of a diffeomorphism invariant theory that
differs from the one used in the loop quantum gravity literature. As
we tried to argue it has some favourable properties like greater
regularity of the representation leading to the existence of
derivatives of the Weyl operator and a Hilbert space of smoother
functions.

Our construction was based on the fact that the string in a flat
target space is a free field theory. However, other exactly solvable
conformal field theories should be treatable in exactly the same fashion
leading to diffeomorphism invariant interacting theories. However it
remains to be seen if similar constructions can also be found for the
case of higher dimensional gravity. But at least it is demonstrated
that there is a viable alternative to singular representations based on
polymer states.

\ack We benefited a lot from discussions with H. Pfeiffer and
T. Thiemann and received useful comments from K. Peeters and
H. Halvorson. 
The work of RCH is partly supported by EU contract
HPRN-CT-2000-00122. Both authors are supported by a PPARC rolling
grant.

\section*{References}
\bibliographystyle{JHEP}
\bibliography{canonstring}

\end{document}